\documentclass[a4paper]{raa}            
\usepackage{CJK}
\usepackage{graphicx,times}             
\usepackage{natbib}
\usepackage{amssymb,amsmath}
\bibpunct{(}{)}{;}{a}{}{,}

\usepackage[a4paper=true,pagebackref=true]{hyperref}
\hypersetup{colorlinks = true, linkcolor = blue, anchorcolor = red, citecolor = blue, filecolor = red, pagecolor = red, urlcolor = red}

\newcommand{\bdv}[1]{\mbox{\boldmath$#1$}}
\def\au{{\rm AU}}
\def\l2{{\rm L2}}
\def\leo{{\rm Leo}}
\def\eff{{\rm eff}}
\def\kms{{\rm km\,s}^{-1}}

\def\te{{t_{\rm E}}}
\def\pie{{\pi_{\rm E}}}
\def\source{{\rm S}}
\def\fbase{{F_{\rm base}}}

\begin{document}
\begin{CJK*}{UTF8}{gbsn}

   \title{Measuring microlensing parallax via simultaneous observations from Chinese Space Station Telescope and \emph{Roman} Telescope}

   \volnopage{Vol.0 (20xx) No.0, 000--000}      
   \setcounter{page}{1}          

   \author{Shi Yan (颜实)
      \inst{1, 2}
   \and Wei Zhu (祝伟)
      \inst{1, 3}
   }

   \institute{
    Department of Astronomy, Tsinghua University, Beijing 100084, China; {\it weizhu@tsinghua.edu.cn} \\
    \and
    School of Physics, Nankai University, Tianjin 300071, China \\
        \and
             Canadian Institute for Theoretical Astrophysics, University of Toronto, Toronto, ON M5S 3H8, Canada \\
\vs\no
   {\small Received~~20xx month day; accepted~~20xx~~month day}}

\abstract{Simultaneous observations from two spatially well-separated telescopes can lead to the measurements of the microlensing parallax parameter, an important quantity toward the determinations of the lens mass. The separation between Earth and Sun-Earth L2 point, $\sim0.01$ AU, is ideal for parallax measurements of short and ultra-short ($\sim$1\,hr to 10\,days) microlensing events, which are candidates of free-floating planet (FFP) events. In this work, we study the potential of doing so in the context of two proposed space-based missions, the Chinese Space Station Telescope (CSST) in a Leo orbit and the Nancy Grace Roman Space Telescope (\emph{Roman}) at L2. We show that the joint observations of the two can directly measure the microlensing parallax of nearly all FFP events with timescales $t_{\rm E}\lesssim$ 10\,days as well as planetary (and stellar binary) events that show caustic crossing features. The potential of using CSST alone in measuring microlensing parallax is also discussed.
}

   \authorrunning{Shi Yan \& Wei Zhu }            
   \titlerunning{Microlensing parallax from L2 \& Leo}  

   \maketitle

%
%
\section{Introduction} \label{sect:intro}

Gravitational microlensing is powerful in detecting cold planets, including those beyond the water snow line and those that are unbound to any stars \citep[e.g.,][]{Mao:1991,GouldLoeb:1992,Sumi:2011}. Statistical studies have found that cold planets are abundant \citep{Gould:2010,Suzuki:2016}, and that unbound planets may be as common as main-sequence stars in the Galaxy \citep{Mroz:2017,Mroz:2019}. Observational constraints on the unbound planets, or free-floating planets (FFPs), can provide important constraints on the formation and evolution of planetary systems (see \citealt{ZhuDong:2021} for a recent review).

The standard microlensing technique usually does not yield measurements of the lens mass. This is because the microlensing timescale, $t_{\rm E}$, involves multiple physical parameters
\begin{equation}
t_{\rm E} \equiv \frac{\theta_{\rm E}}{\mu_{\rm rel}}.
\end{equation}
Here $\mu_{\rm rel}$ is the relative proper motion between the lens and the source, and $\theta_{\rm E}$ is the angular Einstein radius
\begin{equation}
\theta_{\rm E} \equiv \sqrt{\kappa M_{\rm L} \pi_{\rm rel}} ,
\end{equation}
where $\kappa \equiv 4 G (c^{2} \mathrm{AU})^{-1} \approx 8.14\,\mathrm{mas}\,M_{\odot}^{-1}$ is a constant, $M_{\rm L}$ is the mass of the lens, and $\pi_{\rm rel}={\rm AU}(D_{\rm L}^{-1}-D_{\rm S}^{-1})$ is the lens--source relative parallax, with $D_{\rm L}$ and $D_{\rm S}$ the distances to the lens and the source, respectively \citep{Gould:2000}.

The lens mass can be measured or directly constrained if two of the three physical quantities---angular Einstein radius, microlensing parallax, and lens flux---are measured \citep[e.g.,][]{Yee:2015}.
The microlensing parallax is the lens--source relative parallax scaled by the angular Einstein radius \citep{Gould:1992}
\begin{equation}
\pi_{\rm E} \equiv \frac{\pi_{\rm rel}}{\theta_{\rm E}} .
\end{equation}
For faint microlenses such as FFPs, the only way to directly measure the mass is therefore the combination of $\theta_{\rm E}$ and $\pi_{\rm E}$. The microlensing parallax is of particular importance, as it alone can already yield strong constraints on the lens mass under general assumptions of the lens kinematics \citep{Han:1995,Zhu:2017a}.

Many methods have been proposed to measure the microlensing parallax in general microlensing events (e.g., \citealt{Gould:1992,Gould:1994}; see a brief summary in \citealt{Zhu:2015}). In the context of FFP events, the ideal way is to obtain observations from two observatories that are separated by a large fraction of the projected Einstein radius
\begin{equation}
\tilde{r}_{\rm E} \equiv \frac{\rm AU}{\pi_{\rm E}} = 0.016 \left(\frac{M_{\rm L}}{M_\oplus}\right)^{1/2} \left(\frac{\pi_{\rm rel}}{0.1\,{\rm mas}}\right)^{-1/2} \,{\rm AU}.
\end{equation}
Here the normalization has been chosen for an Earth-mass object at a typical microlensing distance ($D_{\rm L}\approx 4\,$kpc). A combination of telescopes at Earth and Sun--Earth L2 point, with a maximum projected separation of 0.01\,AU, is therefore ideal for the detection of microlensing parallax for free-floating planetary-mass objects.

Studies have looked into the feasibility of measuring microlensing parallax from Earth and L2. The \emph{Roman} Space Telescope, previously known as WFIRST, is going to conduct multiple microlensing campaigns from L2 \citep{Spergel:2015}. With 2.4\,m aperture and 0.3\,deg$^2$ field of view, \emph{Roman} is expected to detect over one thousand bound planets and hundreds of free-floating planets out of its $5\times72\,$day microlensing campaigns \citep{Penny:2019,Johnson:2020}. The potential of utilizing a \emph{Roman}-like telescope for microlensing parallax observations has long been realized \citep[e.g.,][]{Gould:2003,Han:2004,Yee:2013,Gould:2021}. In particular, \citet{Zhu:2016} and \citet{Street:2018} studied the use of ground-based telescopes, either the Korean Microlensing Telescope Network \citep[KMTNet,][]{Kim:2016} or the \emph{Rubin} Telescope (previously known as LSST) to augment the microlensing parallax measurements of \emph{Roman}. See also \citet{Bachelet:2019} and \citet{Ban:2020} for the feasibility of using two L2 satellites (i.e., \emph{Euclid} and \emph{Roman}) to determine microlensing parallax parameters.

The Chinese Space Station Telescope (CSST) is a planned mission currently scheduled to launch in late 2023 and start scientific observations in 2024 \citep{Zhan:2011,Zhan:2018,Cao:2018}. CSST has an aperture of 2\,m and a field of view of 1.1\,deg$^2$, and it will be in a $\sim400\,$km low-Earth orbit (Leo) with an orbital period around 95\,min. The primary science goal of CSST is to understand the nature of dark matter and dark energy using a number of cosmological probes, such as galaxy clusterings as well as weak and strong gravitational lensings \citep[e.g.,][]{Gong:2019,Zhang:2019}. The high resolution and large field of view of CSST make it also a wonderful mission to conduct space-based microlensing surveys \citep{Gould:2009}. Similar to those by \emph{Euclid} and \emph{Roman} \citep{Penny:2013,Penny:2019,Johnson:2020}, the microlensing survey by CSST will have the sensitivity to low-mass planets at both wide and unbound orbits.

This paper discusses the potential of measuring microlensing parallax by CSST alone and by combining CSST at Leo and a \emph{Roman}-like telescope at L2. We show in Section~\ref{sec:pspl} that such a joint program can measure microlensing parallax for the majority of short (and ultra-short) timescale microlensing events. Section~\ref{sec:bi} explains the potential of CSST and \emph{Roman} in measuring parallax and thus full lens solution of caustic-crossing binaries (including planetary events). A brief discussion is given in Section~\ref{sec:discuss}.

\section{Measuring the microlensing parallax of single-lens events} \label{sec:pspl}

The microlensing light curve arising from a point source and a single lens is given by \citep{Paczynski:1986}
\begin{equation} \label{flux_def}
F (t) = F_\source [A(t)-1] + \fbase,\quad
A(t) = \frac{u^2+2}{u\sqrt{u^2+4}} .
\end{equation}
Here $F_\source$ is the source flux at baseline, $\fbase$ is the total flux (source \& possible blend) at baseline, and $u$ is the distance between the source and the lens at a given time $t$ normalized to the Einstein radius. Because of the microlensing parallax effect, telescopes at different locations (namely L2 and Leo) see different values of $u$
\begin{equation} \label{eqn:u}
u^2_\l2 = \frac{(t-t_{0,\l2})^2+t_{\eff,\l2}^2}{\te^2}; \quad
u^2_\leo= \frac{(t-t_{0,\leo})^2+t_{\eff,\leo}^2}{\te^2} .
\end{equation}
Here $t_0$ is the peak time and $\te$ is the event timescale. We use the effective timescale $t_\eff \equiv u_0 \te$ rather than the impact parameter $u_0$ in describing the event evolution, because $t_\eff$ is almost always better constrained in observations than is $u_0$.

The microlensing parallax measured from two well separated observatories can be approximated as \citep{Refsdal:1966,Gould:1994}
\begin{equation} \label{eqn:parallax_vector}
\bdv\pie=(\pie_\parallel,\, \pie_\perp)=\frac{\au}{D_{\perp}}\left(\frac{\Delta t_{0}}{\te}, \Delta u_{0}\right),
\end{equation}
where $\pie_{\parallel}$ and $\pie_{\perp}$ are the components parallel and perpendicular to the source trajectory, respectively. Here $D_{\perp}$ is the projected separation between the two observatories, $\te$ is the event timescale, and $\Delta t_{0}\equiv t_{0, \leo}-t_{0, \l2}$ and $\Delta u_{0} \equiv u_{0, \leo}-u_{0, \l2}$ are the differences in peak times and impact parameters as seen from two observatories, respectively. 

By writing the parallax vector in the form of Equation~(\ref{eqn:parallax_vector}) we have assumed that (a) the event timescale $\te$ is (effectively) the same at two locations and that (b) both $\te$ and the projected separation $D_\perp$ are invariant in the process of the event. The latter is a reasonable assumption for short (less than a few days) timescale events, which are the focus of this section. Regarding the former, the difference in $\te$ arises from the relative motion between the two telescopes, which we separate into two parts. The relative motion between \emph{Roman} and Earth ($<1\,\kms$) is very small compared to the lens-source projected motion, which is typically $\tilde{v}_{\rm hel}\gtrsim200\,\kms$ (see Appendix B of \citealt{Zhu:2016}). The relative motion between CSST and Earth varies within an orbital period of CSST ($T\approx90\,$min) and caps at $\sim8\,\kms$. For events with $\te\gg T$, the time-averaged relative velocity is negligible compared to the transverse velocity $\tilde{v}_{\rm hel}$. For events with $\te$ comparable to $T$, the relative motion between CSST and Earth is in fact useful in further breaking the standard four-fold degeneracy in two-location observations \citep{Refsdal:1966,Gould:1994,Zhu:2017b}, as in such cases CSST can be effectively treated as more than one static telescopes. We will further discuss the use of this effect in Section~\ref{sec:csst-only}.

While $\bdv{\pie}$ is  an  important  quantity that is frequently used in the literature, a better parameter that is more directly constrained in observations and better connected to the lens kinematics is \citep{Dong:2007,Zhu:2016}
\begin{equation} \label{eqn:lambda}
\bdv{\Lambda} \equiv \frac{\bdv{\pie} \te}{\au} = \frac{1}{\tilde{v}_{\rm hel}} = \frac{1}{D_\perp} (\Delta t_0, \Delta t_\eff) .
\end{equation}
We therefore use this parameter to quantify the detectability of the microlensing parallax effect in such a Leo+L2 configuration.

\begin{figure}
    \centering
    \includegraphics[width=0.9\textwidth]{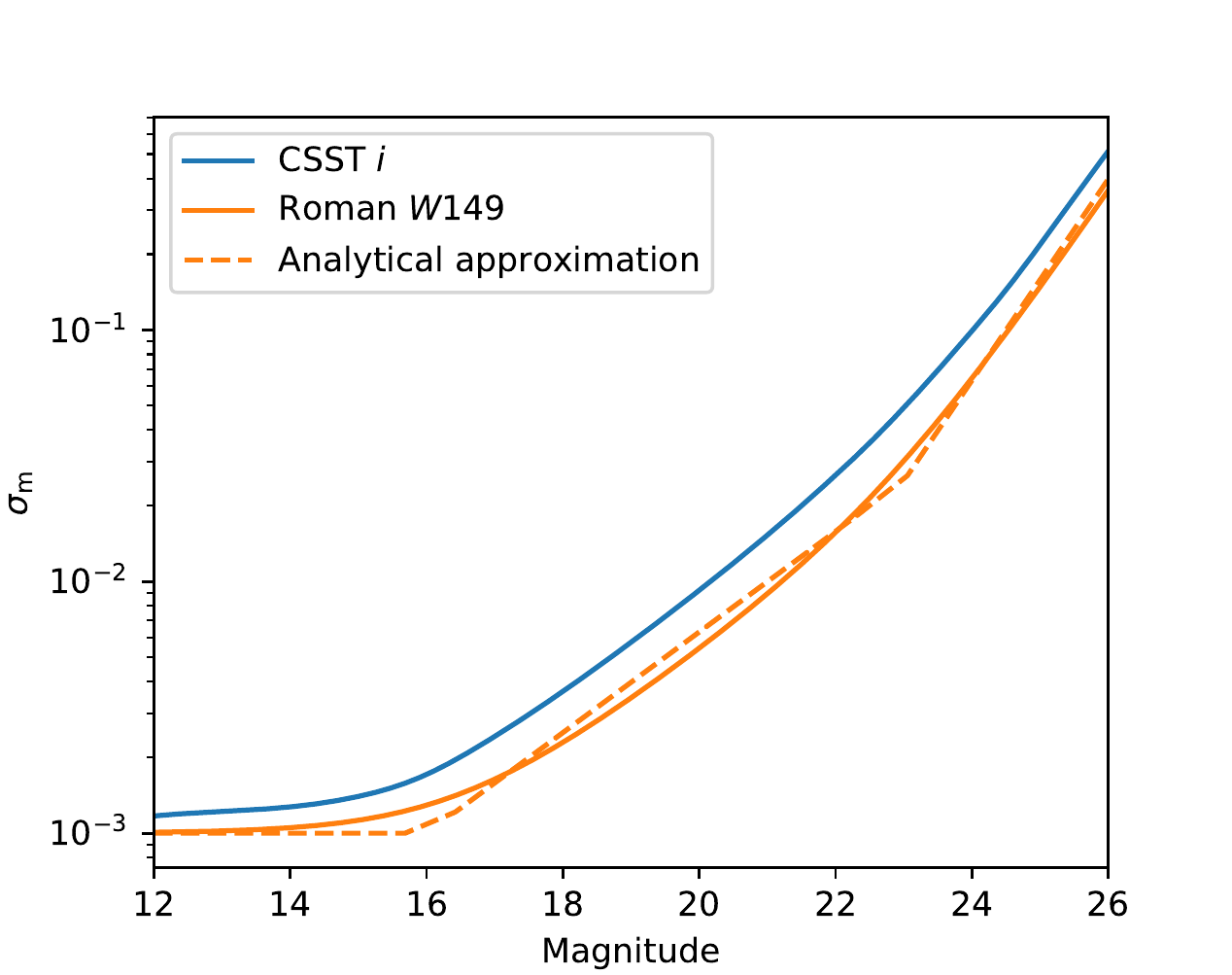}
    \caption{The expected noise curves of the CSST and \emph{Roman} microlensing surveys. The dashed curve shows our analytical approximation of the \emph{Roman} noise curve, which is used in Appendix~\ref{sec:Appendix1} to derive the scaling relations.}
    \label{fig:noise-curves}
\end{figure}

\begin{figure}
    \centering
    \includegraphics [ width=0.9\textwidth ] {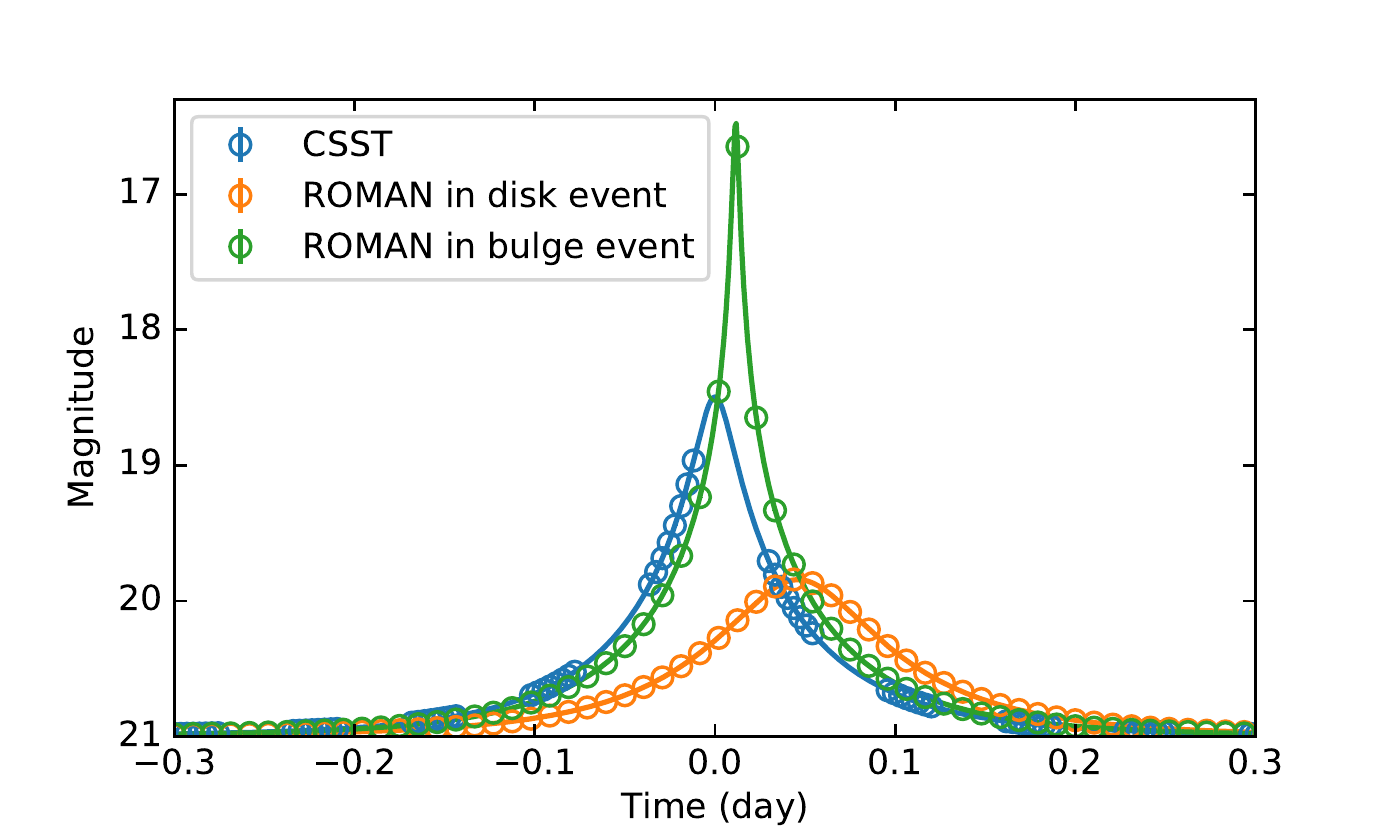}
    \caption{Example light curves of typical disk and bulge events with a timescale $t_{\rm E}=0.1$ days. The blue curve shows the light curve seen by CSST at Leo, for which we have set the impact parameter $u_{0,\rm Leo}=0.1$ and a source baseline flux $F_{\rm S,Leo}=1$ (corresponding to 21 mag). The orange and green curves are what \emph{Roman} would see. We have adopted $\pi_{\rm rel} = 0.12$ mas and $\mu_{\rm rel} = 7$ mas/yr for the disk event, and $\pi_{\rm rel} = 0.02$ mas and $\mu_{\rm rel} = 4$ mas/yr for the bulge event.}
    \label{fig:lcs}
\end{figure}

\begin{figure}
    \centering
    \includegraphics [ width=1.0\textwidth ] {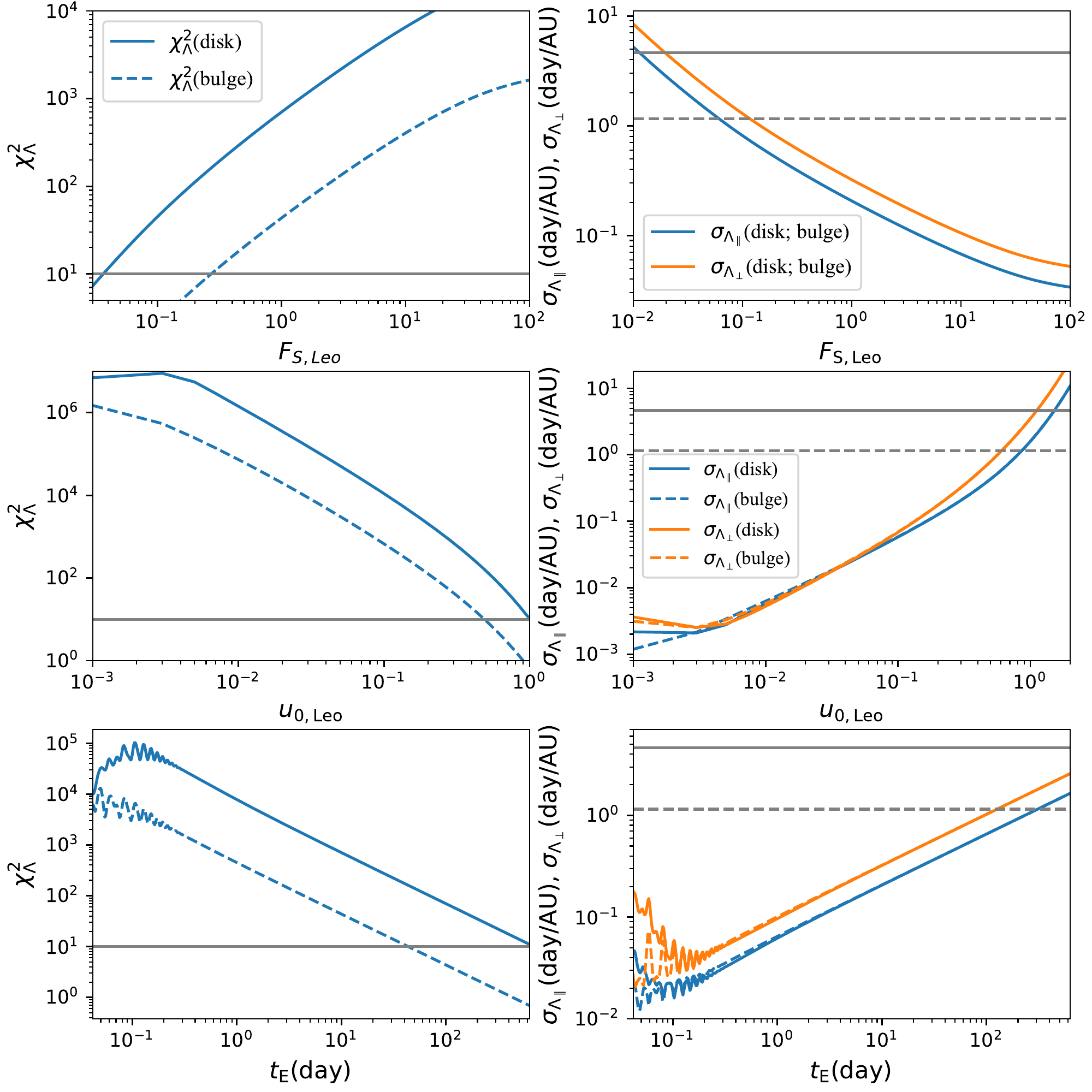}
    \caption{The detection significance of the parallax effect as functions of various parameters. From top to bottom, they are source baseline flux $F_{\rm S,Leo}$, impact parameter $u_{0,\rm Leo}$, and event timescale $t_{\rm E}$, respectively. Panels on the left show the $\chi^2$ significance between models with and without parallax effect, whereas panels on the right show the uncertainties of the kinematic parallax parameters $\Lambda_{\perp}$ and $\Lambda_{\parallel}$. In each panel, the solid curve assumes typical disk lenses (with $\pi_{rel} = 0.125$ mas, $\mu_{rel} = 7$ mas/yr, and thus $\Lambda = 6.52$), and the dashed curve assumes typical bulge lenses (with $\pi_{rel} = 0.018$ mas, $\mu_{rel} = 4$ mas/yr, and thus $\Lambda = 1.63$). In the left panels, the gray horizontal lines represent $\chi^2 = 10$, above which the parallax effect is considered to be detected. In the right panels, the gray horizontal lines denote the values of $\Lambda$. Our default parameters are $F_{\rm S,Leo}=1$, $u_{0,\rm Leo}=0.3$, $t_{\rm E}=10$ days and color parameter $c=1$.}
    \label{fig:main}
\end{figure}

\begin{figure}
    \centering
    \includegraphics [ width=0.8\textwidth ] {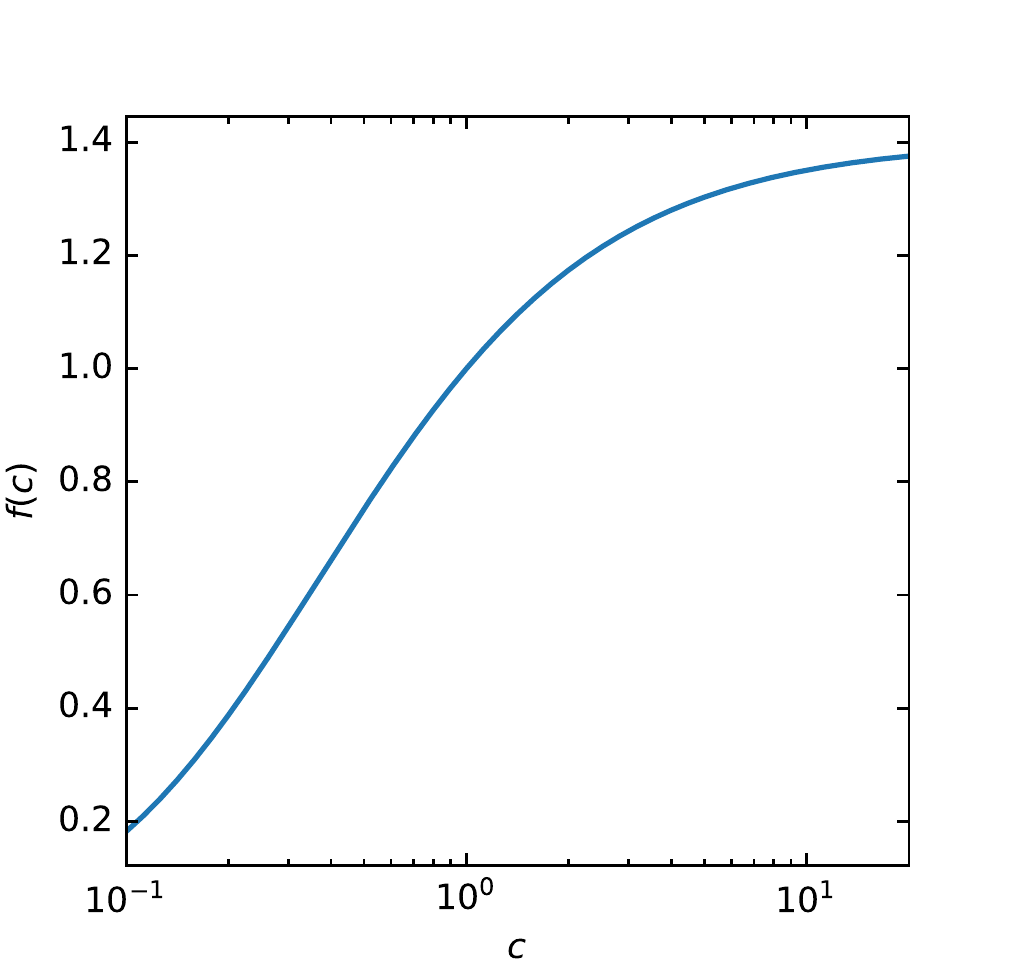}
    \caption{The dependence of the parallax detectability on the color parameter $c$ (defined by Equation~\ref{eqn:color}). The $y$-axis shows the relative enhancement in the parallax detectability, quantified by $\chi^2_{\bdv{\Lambda}}$, of a source with a color parameter $c$ compared to the case with $c=1$.}
    \label{fig:b}
\end{figure}

We use the Fisher matrix analysis to evaluate the detectability of the parallax effect. For the \emph{Roman} survey, we use 15\,min cadence and the expected photometric performance out of the 53\,s exposure from \citet{Penny:2019}. For the CSST survey, we assume a duty cycle of 40\% and 8 observations per orbit (corresponding to $\sim$5\,min cadence). The photometric performance of the CSST microlensing survey is derived from the online CSST ETC simulator
\footnote{\url{http://etc.csst-sc.cn/ETC-nao/etc.jsp}}
. We have assumed high zodiacal light and average Earth shine light for the background contamination and an extinction law of $E(B-V)=0.94$. A typical M0V star in the bulge is chosen as the microlensing source. With an exposure time of 60\,s and a systematic floor of $0.001$ mag, we can then simulate the noise curve of the CSST microlensing survey. The result is shown in Figure~\ref{fig:noise-curves} together with the noise curve of \emph{Roman} from \citet{Penny:2019}.

The parameters used to model the combined data set are
\begin{equation}
    \bdv{\theta}=(\te,\,t_{0,\l2},\,t_{\eff,\l2},\,F_{\source,\leo},\,t_{0,\leo},\,t_{\eff,\leo},\,c) .
\end{equation}
Here $F_{\rm S,Leo}$ is the source flux at baseline in the bandpass of CSST at Leo, and we renormalize the flux values such that $F_{\rm S,Leo}=1$ corresponds to a 21 magnitude star. For the other observatory, we use the parameter $c$ to quantify the ratio of source fluxes in the bandpasses of the two telescopes and it is related to the observed color of the source $m_\leo-m_\l2$
\begin{equation} \label{eqn:color}
    c \equiv \frac{F_{\source,\l2}}{F_{\source,\leo}} = 10^{(m_{\leo}-m_{\l2})/2.5}.
\end{equation}
We do not specify the bandpasses of the two spapce-based surveys, although for practice reasons we will use $H$- and $I$-band for observations from L2 and Leo, respectively. Using the stellar colors from \citet{Pecaut:2013} and assuming an extinction of $A_I=1.5$ and $E(I-H)\approx1$ \citep{Gonzalez:2012,Nataf:2013}, one finds that a Sun-like star in the bulge will have $m_\leo \approx 20$, $m_\leo-m_\l2\approx 1.8$, and $c\approx 5.2$. Early M-dwarfs, which will probably be the primary sources of the events simultaneously observed by CSST and \emph{Roman}, will have $m_\leo \approx 23$, $m_\leo-m_\l2\approx 2.7$ and thus $c \approx 12$.

Throughout this paper we consider two types of microlensing events: a typical disk event with lens-source relative parallax $\pi_{\rm rel}=0.12\,$mas (corresponding to a lens at $\approx4\,$kpc) and a relative proper motion $\mu_{\rm rel}=7\,{\rm mas\,yr^{-1}}$; and a typical bulge event with $\pi_{\rm rel}=0.02\,$mas (corresponding to a lens at $\approx7\,$kpc) and $\mu_{\rm rel}=4\,{\rm mas\,yr^{-1}}$. These yield typical transverse velocities \citep{Zhu:2016}
\begin{equation}
\tilde{v} = \frac{1}{\Lambda} = \left\{
\begin{array}{ll}
280\,{\rm km\,s^{-1}}, & {\rm Disk~events} \\
1000\,{\rm km\,s^{-1}}, & {\rm Bulge~events}
\end{array} \right.
\end{equation}
Figure~\ref{fig:lcs} shows example light curves of typical disk and bulge microlensing events with $t_{\rm E}=0.1\,$d, $u_{\rm 0,Leo}=0.1$, and a source baseline flux $F_{\rm S,Leo}=1$ (corresponding to a baseline magnitude of 21).

The Fisher matrix $\mathcal{F}_{ij}$ is then given by
\begin{equation}
\mathcal{F}_{ij} \equiv \sum_{\{t_l\}} \frac{1}{\sigma_F^2(t_l)} \frac{\partial F_\l2(t_l)}{\partial \theta_i} \frac{\partial F_\l2 (t_l)}{\partial \theta_j} + \sum_{\{t_m\}} \frac{1}{\sigma_F^2(t_m)} \frac{\partial F_\leo(t_m)}{\partial \theta_i} \frac{\partial F_\leo(t_m)}{\partial \theta_j} .
\label{fisher}
\end{equation}
Here $\{t_l\}$ and $\{t_m\}$ are the time series of \emph{Roman} telescope and CSST, respectively, and $\sigma_F$ represents the uncertainty of the measured flux. We then compute the covariance matrix of the vector $\bdv{\Lambda}$, $\Sigma_\Lambda$, based on the relation between $\bdv{\Lambda}$ and the direct observables (equation~\ref{eqn:lambda}) The detectability of the parallax effect is quantified by
\begin{equation} \label{eqn:chi2_definition}
    \chi^2_{\bdv{\Lambda}} = (\bdv{\Lambda}-\bdv{0}) \Sigma_\Lambda^{-1} (\bdv{\Lambda}-\bdv{0})^T ,
\end{equation}
where $\bdv{0}$ denotes the zero-parallax case.

For any chosen set of microlensing parameters we apply the above Fisher matrix analysis to numerically evaluate the detectability of the parallax effect. One example output is shown in Figure~\ref{fig:main}, for which we have assumed a maximum projected separation between Leo and L2 $D_\perp=0.01$ AU.

To gain theoretical insights, we also derive the following analytical scaling relation in the high-magnification ($u_0\rightarrow0$) regime (see Appendix~\ref{sec:Appendix1} for the detailed derivation)
\begin{equation} \label{eqn:scaling}
\chi_{\bdv\Lambda}^{2}\propto \frac{D_\perp^2 F_{\rm S,Leo}}{t_{\rm E} u_{0,\leo}^2} f(c)
\end{equation}
Here $f(c)\equiv \chi^2_\Lambda(c)/\chi^2_\Lambda(c=1)$ captures the dependence of $\chi^2_\Lambda$ on the color parameter $c$ and is normalized by the value of $\chi^2_\Lambda$ at $c=1$. As shown in Figure~\ref{fig:main}, the above scaling relation matches the numerical results reasonably well and breaks down at both large ($u_0\gtrsim1$) and small $u_0$ values ($u_0\lesssim0.01$). The former is due to the breakdown of the high-magnification regime. The latter is due to the fact that we have assumed $u_{\rm 0,Leo}=u_{\rm 0,L2}$ in the analytical derivation. According to Equation~(\ref{eqn:parallax_vector}), $\Delta u_0 = \pi_{\rm E,\perp} D_\perp$. For typical disk and bulge events, $\Delta u_0\approx0.01$ and $0.001$, respectively, and thus $u_{\rm 0,L2}=u_{\rm 0,Leo}+\Delta u_0 \rightarrow u_{\rm 0,Leo}$ as long as $u_{\rm 0,Leo}\gg0.01$. Additionally, the above scaling typically breaks down at $t_{\rm E}\lesssim 0.1\,$days when the event timescale is comparable to the orbital period of the Leo satellite around Earth.

With Equation~(\ref{eqn:scaling}) and the numerical results from Figure~\ref{fig:main}, we can rewrite the detection significance of the parallax effect in $\chi_{\bdv\Lambda}^2$ as the following expressions
\begin{equation} \label{eqn:chisq}
    \chi_{\bdv\Lambda}^2 = \left\{
    \begin{array}{ll}
        37 \left(\frac{F_{\rm S,Leo}}{1}\right) \left(\frac{t_{\rm E}}{10\,\rm days}\right)^{-1} \left(\frac{u_{0,\rm Leo}}{0.3}\right)^{-2} \left(\frac{D_\perp}{0.01\,\rm AU}\right)^2 f(c) , & {\rm Bulge~event} \\
        274 \left(\frac{F_{\rm S,Leo}}{1}\right) \left(\frac{t_{\rm E}}{10\,\rm days}\right)^{-1} \left(\frac{u_{0,\rm Leo}}{0.3}\right)^{-2} \left(\frac{D_\perp}{0.01\,\rm AU}\right)^2 f(c) , & {\rm Disk~event} \\
    \end{array}
    \right.
\end{equation}
The above expressions can be readily used to evaluate the detectability of the parallax effect in any given single-lens event. Specifically, with the maximum projected separation of $D_\perp=0.01$ AU between Earth and L2, the combination of CSST and \emph{Roman} should be able to detect the microlensing parallax effect for typical bulge (disk) events with $0.1\lesssim t_{\rm E}/{\rm days}\lesssim 10$ ($0.1\lesssim t_{\rm E}/{\rm days}\lesssim 80$), if the source is 23 mag at baseline and the impact parameter is $\sim0.3$. These values enclose the bulk of the expected FFPs from \emph{Roman}, especially those with low-mass (and thus preferentially terrestrial) lenses \citep[e.g.,][]{Johnson:2020}.

\subsection{Lens mass determinations}

\begin{figure}
    \centering
    \includegraphics[width=0.8\textwidth]{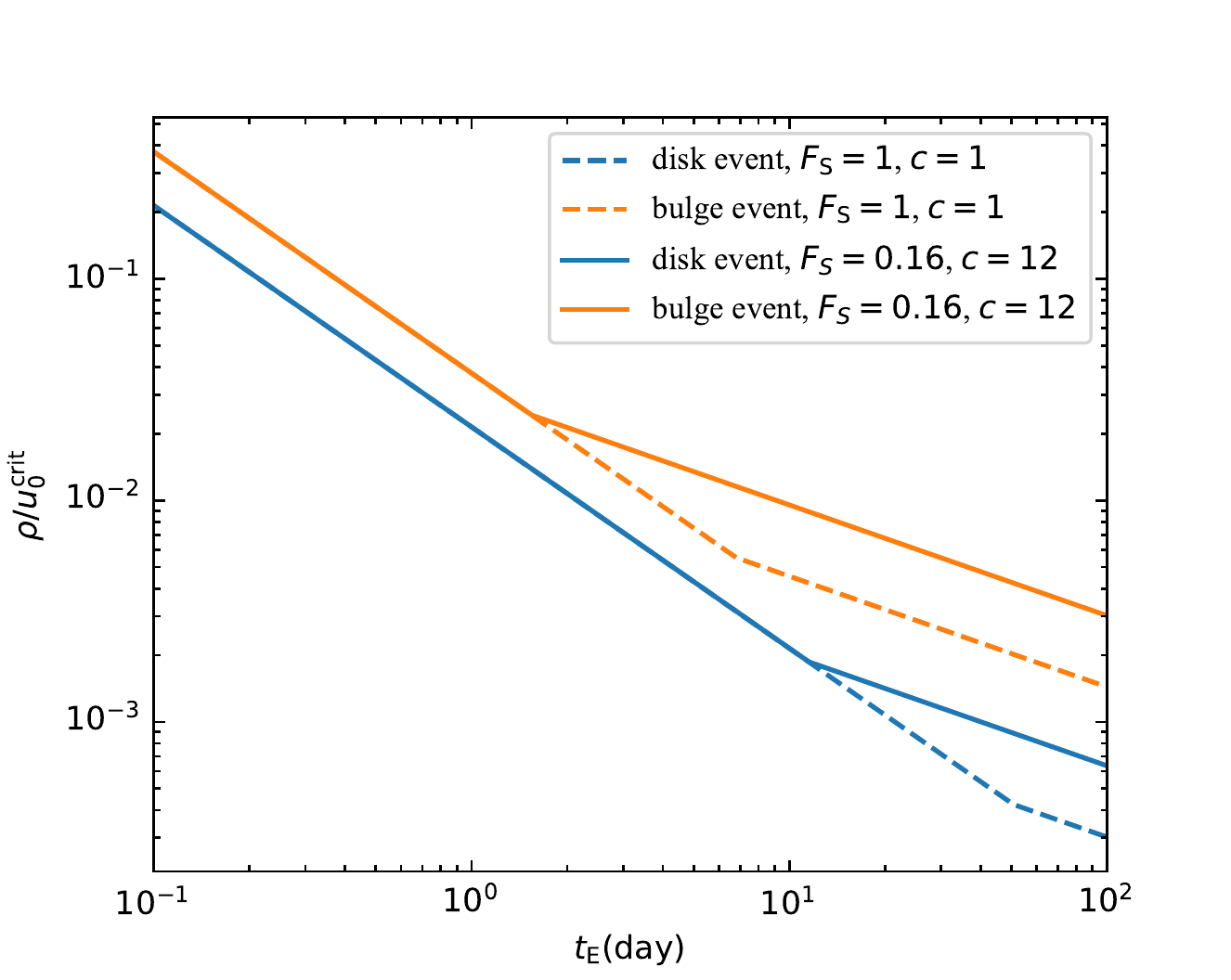}
    \caption{The fraction of microlensing events with full lens mass measurement out of all with microlensing parallax effect, quantified by ratio between the scaled source size $\rho$ and the critical impact parameter for parallax detections $u_0^{\rm crit}$. See the main text for more details. Here blue and orange curves correspond to typical disk and bulge events, respectively. The dashed curves are for our default set of source parameters ($F_{\rm S}=1,~c=1$), whereas the solid curves are for more typical microlensing sources from the \emph{Roman} survey ($F_{\rm S}=0.16,~c=12$).}
    \label{fig:FS}
\end{figure}

Besides the microlensing parallax, another critical paramter toward a lens mass measurement is the angular Einstein radius $\theta_{\rm E}$, which is usually determined from the finite-source effect \citep{Yoo:2004}. For single-lens events, the finite-source effect becomes prominent when the impact parameter $u_0$ becomes comparable or less than the scaled source size $\rho$. The fraction of microlensing events with complete lens mass measurements out of all events with parallax detections can be estimated as
\begin{equation}
    P_{\rm mass} \sim \frac{\rho}{u_0^{\rm crit}} = \frac{\theta_\star}{\mu_{\rm rel} t_{\rm E} u_0^{\rm crit}} .
\end{equation}
Here $\theta_\star=0.3\,\mu$as is the typical value for the source angular size , which corresponding to an early M dwarf with a radius of $0.5\,R_\odot$ in the Galactic Bulge. The quantity $u_0^{\rm crit}$ is the critical impact parameter above which the parallax effect is no longer detectable. For given source and lens properties, it can be derived from the scaling relations in Equation~(\ref{eqn:chisq}) with a certain $\chi^2_{\bdv{\Lambda}}$ threshold for the parallax detection. We use a $\chi^2_{\bdv{\Lambda}}$ threshold of 10 and set an upper limit on $u_0^{\rm crit}$ at 0.7. The latter takes into account the fact that the scaling relations are no longer valid approximations when $u_0$ approaches unity.

Figure~\ref{fig:FS} shows the estimated fraction of events with complete mass measurements out of those with parallax detections. Here we have included the results with typical sources ($F_{\rm S}=0.16$ and $c=12$, corresponding to $m_{\rm Leo}\approx23$ and $m_{\rm Leo}-m_{\rm L2}\approx2.7$) in addition to the results with the default $F_{\rm S}=1$ and $c=1$ combination.
For events with timescales in the range of 0.1--10 days, for which the joint observations of CSST and \emph{Roman} are most valuable, $\gtrsim0.3\%$ ($\gtrsim2\%$) of disk (bulge) events with typical sources and parallax detections should have complete lens mass determinations.
Up to $\sim$40\% of ultra-short-timescale events will show the finite-source effect and thus have complete lens mass determinations. This fraction is largely independent on the source properties. 

For events that do not show the finite-source effect, the lens mass can be inferred from the microlensing parallax alone \citep{Han:1995}. Such mass inferences are statistically accurate at $\sim30\%$ level for typical disk events \citep[e.g.,][]{Yee:2015,Zhu:2017a}, whereas the uncertainty increases to $\gtrsim40\%$ for bulge events \citep{Gould:2021}. Such statistically inferred lens properties are also valuable in constraining the mass distribution of isolated microlenses, especially those with low probabilities of showing the finite-source effect.

\subsection{CSST alone} \label{sec:csst-only}

\begin{figure}
    \centering
    \includegraphics [ width=0.9\textwidth ] {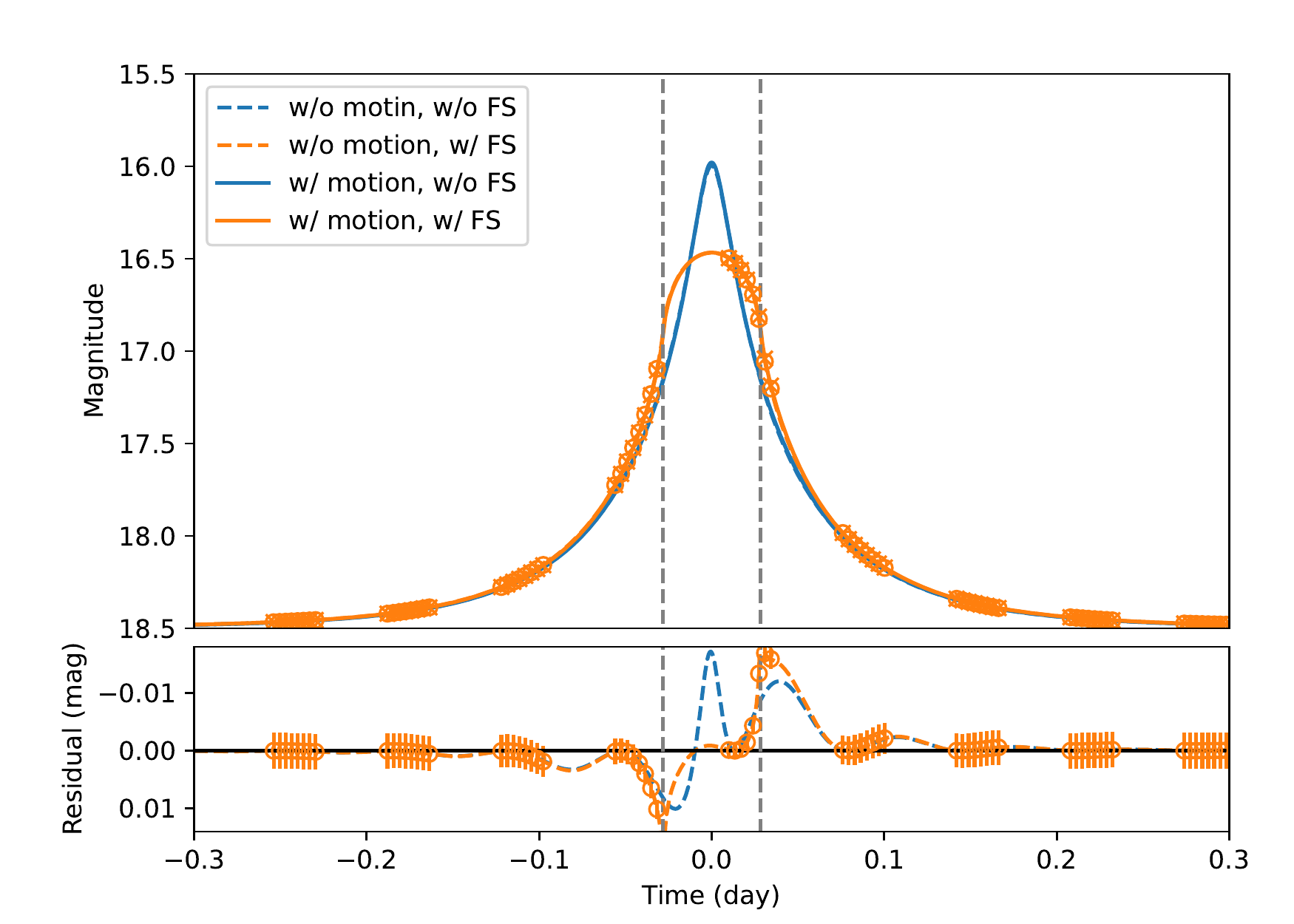}
    \caption{An ultra-short-timescale ($t_{\rm E}=0.1\,$days) event arising from a typical disk lens seen by CSST alone. With an impact parameter of $u_0=0.1$, the event with a scaled source size $\rho=0.3$ shows strong finite-source (FS) effect (orange curves), where the peak region of the light curve deviates substantially from the point source approximation (blue curves). The dashed vertical lines indicate the locations where the lens--source distance $u=\rho$. In each set of curves, the solid one implements the orbital motion of CSST, and the resulting light curve appears slightly but statistically significantly different from the dashed curve. Without the finite-source effect, the $\chi^2$ difference between the dashed and the solid curves is $141$. With the finite-source effect, the $\chi^2$ difference increases to $169$.}
    \label{fig:csst}
\end{figure}

The orbital motion of CSST around Earth will be useful in further resolving the four-fold degeneracy in the single-lens case \citep{Refsdal:1966,Gould:1994,Zhu:2017b} as well as detecting the parallax effect by CSST alone. An example light curve to demonstrate the use of the orbital motion effect of CSST is shown in Figure~\ref{fig:csst}.

For simplicity, we will treat the orbital motion effect of CSST as if two satellites were simultaneously observing with a separation of an Earth diameter (i.e., $R_\perp=2R_\oplus$). With this approximation, the result from the CSST+\emph{Roman} analysis (Equation~\ref{eqn:chisq}) is directly applicable
\begin{equation} \label{eqn:csst}
    \chi_{\bdv\Lambda, \rm CSST}^2 \approx \left\{
    \begin{array}{ll}
        3.7 \left(\frac{F_{\rm S,Leo}}{1}\right) \left(\frac{t_{\rm E}}{0.1\,\rm days}\right)^{-1} \left(\frac{u_{0,\rm Leo}}{0.1}\right)^{-2} \left(\frac{D_\perp}{2\,R_\oplus}\right)^2 , & {\rm Bulge~event} \\
        27 \left(\frac{F_{\rm S,Leo}}{1}\right) \left(\frac{t_{\rm E}}{0.1\,\rm days}\right)^{-1} \left(\frac{u_{0,\rm Leo}}{0.1}\right)^{-2} \left(\frac{D_\perp}{2\,R_\oplus}\right)^2 , & {\rm Disk~event} \\
    \end{array}
    \right.
\end{equation}
Note that these scaling relations do not take into account the finite-source effect. The point-source example shown in Figure~\ref{fig:csst} has $\chi^2_{\bdv{\Lambda}}=141$, which for the chosen parameters is a factor of $\sim$2 smaller than what the above scaling relation yields. This difference probabaly comes from the fact that the real orbit of CSST has an effective projected separation smaller than $2\,R_\oplus$.

Ultra-short-timescale events with relatively small impact parameters almost always show the finite-source effect, so it is necessary to take into account the impact of this additional effect on the detectability of the parallax effect. The finite-source effect works in two different ways: on the one hand, it reduces the magnification when the source--lens relative distance $u \ll \rho$; on the other hand, the finite-source effect increases the magnification at $u\sim\rho$ (\citealt{Gould:1994b}, see also \citealt{Yoo:2004,Chung:2017}). Whether the finite-source magnification is less or more sensitive to the variation of distance $u$ than is the point-source magnification depends on the ratio between $u$ and $\rho$. Consequently, the detectability of the parallax effect can be either enhanced (when there are more data points with $u \sim \rho$) or reduced (when there are more data points with $u \ll \rho$). For example, the finite-source example shown in Figure~\ref{fig:csst} has $\chi^2_{\bdv{\Lambda}}=161$, slightly higher than the value of the point-source case. Considering that the source should be relatively bright, here we have assumed a Sun-like star in the bulge as the microlensing source.

A more precise and comprehensive evaluation of the parallax detectability in the case of CSST alone will require additional knowledge like the position of Earth and the inclination of CSST orbit, and thus we do not attempt to derive in the present work. In short, CSST alone should be able to detect the parallax effect of ultra-short-timescale events which have relatively high magnifications and bright sources (see also \citealt{Mogavero:2016}). For such events, the joint observations of CSST and \emph{Roman} can also be used to break the four-fold degeneracy that is generic to single-lens microlensing events.

\section{Measuring microlensing parallax of caustic-crossing binary events} \label{sec:bi}

\begin{figure}
    \centering
    \includegraphics [ width=1\textwidth ] {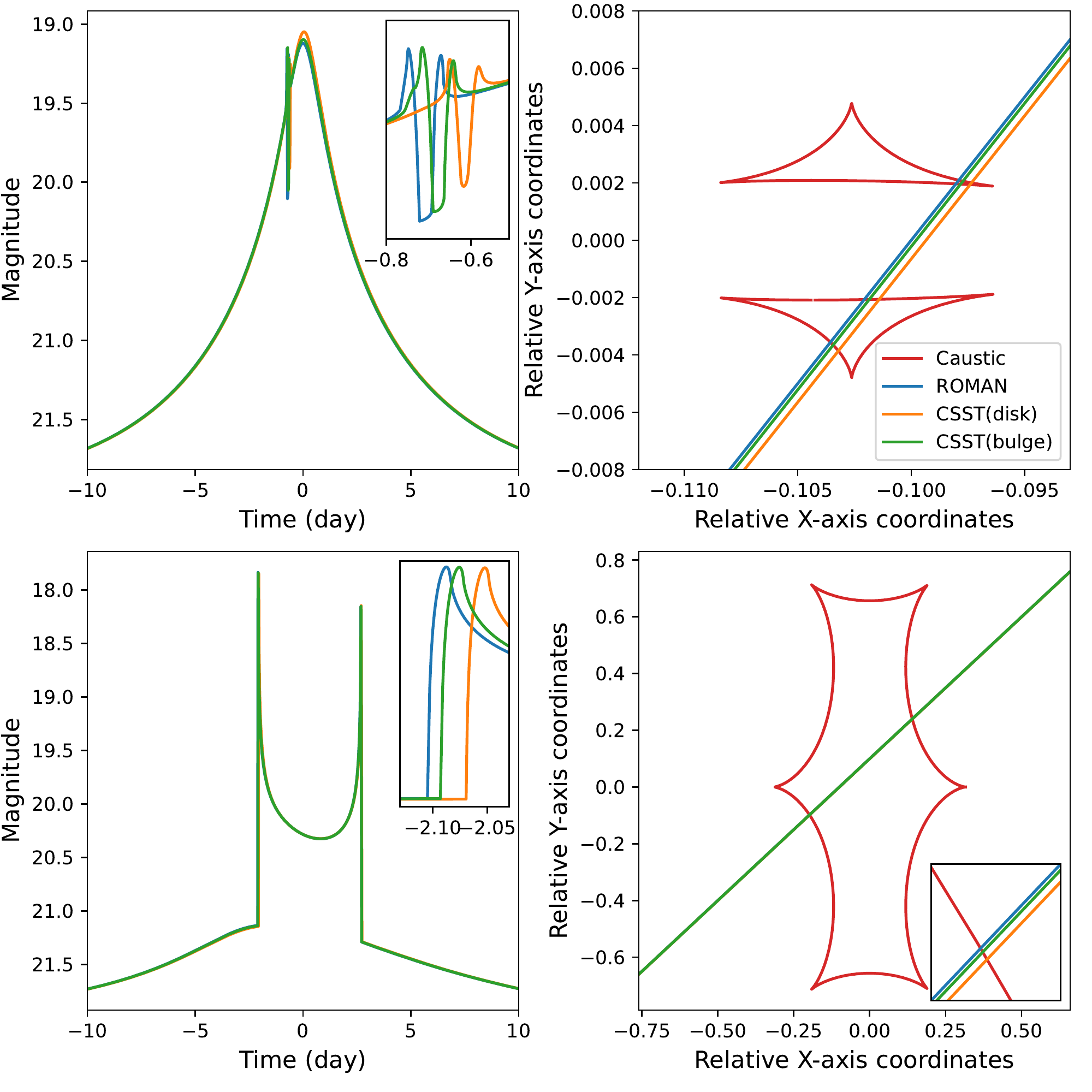}
    \caption{Typical planetary (upper panels, with mass ratio $q=10^{-5}$) and binary (lower panels, with $q=1$) microlensing events with caustic-crossing features seen by CSST at Leo and \emph{Roman} at L2. The left panels show the expected light curves and the right panels show the caustic and lens-source relative trajectories. We have assumed a timescale $t_{\rm E}=10\,$days and scaled source size $\rho=10^{-3}$ in both cases. Although the light curves seen by CSST and \emph{Roman} look overall very similar, the caustic-crossing features show statistically significant differences that can be used to measure the parallax effect.}
    \label{fig:binary}
\end{figure}

For binary and planetary events with caustic-crossing features, which constitute roughly half of all binary and planetary events in such well sampled observing campaigns \citep{Zhu:2014}, the joint observations of CSST and \emph{Roman} can also measure the microlensing parallax and thus fully determine the mass and distance of the lens system. The caustic-crossing features are smoothed on the scales of the source size, and thus two observers separated by roughly the (projected) source size should see caustic-crossing features with measurable time offsets. For typical microlensing events in the CSST and \emph{Roman} campaigns, the source stars are sub-solar in size \citep[e.g.,][]{Penny:2019}, and thus the Leo--L2 separation, $\sim0.01\,{\rm AU}=2\,R_\odot$, is ideal for the microlensing parallax observations regardless of the event timescale (see also \citealt{Gould:2021}). Figure~\ref{fig:binary} illustrates two example light curves, one of planetary nature with mass ratio $q=10^{-5}$ and the other of binary nature with $q=1$. In both examples, the caustic-crossing features are significantly offsetted in the light curves of CSST and \emph{Roman}.

Besides the Leo--L2 separation, other factors also affect the detectability of the microlensing parallax. The determination of the full two-dimensional microlensing parallax requires that each satellite should capture (at least) two caustic crossing features. Otherwise there will be a continuous degeneracy in the parallax vector \citep{HardyWalker:1995,Graff:2002}. This is not rare in reality. For example, some caustic-crossing events have effectively only one caustic-crossing feature \citep[e.g.,][]{Shvartzvald:2015}, or the observations of one or both satellites may fail to capture some of the caustic-crossing features \citep[e.g.,][]{Zhu:2015}. This latter scenario is probably going to be the dominant factor, as CSST at Leo has a duty cycle of $\sim$30--40\% due to (mostly) Earth's umbra.

Caustic-crossing events may have their parallax further constrained by the orbital motions of CSST or \emph{Roman} \citep{Honma:1999}. Such a complementary constraint will be especially useful for caustic-crossing events with continuous degeneracies. We refer to \citet{Zhu:2016} for more discussion.

\section{Discussion} \label{sec:discuss}

This work studies the potential of simultaneous observations by CSST and \emph{Roman} in detecting the microlensing parallax effect. We have derived the scaling relations that quantify the detectability of the parallax effect by the two satellites together and by CSST alone. Our calculations show that CSST and \emph{Roman} together can measure the microlensing parallax for short and ultra-short timescale ($\sim$0.1--10\,days) microlensing events, which roughly correspond to brown dwarf and lower-mass lenses. CSST alone has the capability to detect the microlensing parallax for ultra-short timescale events with relatively high magnifications and bright sources. Furthermore, The separation between CSST and \emph{Roman} is ideal for microlensing parallax measurements of caustic-crossing events. With the angular Einstein ring radius measurements from the caustic-crossing features \citep{Yoo:2004}, these parallax measurements will lead to direct determinations of the mass and the distance of the lens system.

Although the combination of Leo and L2 is not ideal to detect the parallax effect of typical Galactic microlensing events, which usually require a projected separation between observatories of several AU \citep{Refsdal:1966,Gould:1994}, it is perhaps the best approach to complete the mass measurements (or inferences) of Galactic microlenses. A space-based telescope with high spatial resolution like CSST and \emph{Roman} can resolve individual bulge stars and thus constrain the properties of luminous lenses through the lens flux method \citep[e.g.,][]{Bennett:2007}. The microlensing campaign by CSST or \emph{Roman} alone can therefore measure/constrain the mass function of (most) stellar lenses. However, as the lens flux method does not work for sub-stellar microlenses such as brown dwarfs and FFPs, the microlensing parallax method becomes the only way to constrain the lens properties. Therefore, a joint program that can detect the parallax effect of short and ultra-short timescale microlensing events can complement the lens mass measurements/constraints of the microlensing compaigns of individual missions.
 
Finally, it is also worth pointing out that although our results have assumed the configuration of CSST and \emph{Roman}, they are generally applicable to any combination of satellites at Leo and L2.

\begin{acknowledgements}
We would like to thank the anonymous referee for comments and suggestions that have improved the paper.
We acknowledge the science research grants from the China Manned Space Project with No.\ CMS-CSST-2021-A11.
\end{acknowledgements}

\appendix  

\section{Fisher matrix analysis: analytics} \label{sec:Appendix1}

The scaling relation (Equation~\ref{eqn:chisq}) can be derived analytically in the high-magnification regime, where the light curve can be modeled with only three parameters $(t_0, t_{\rm eff}, F_{\rm peak})$ \citep{Gould:1996}
\begin{equation}
F(t)=F_{\rm peak} Q(t); \quad Q(t)=\left(\tau_{\rm eff}^{2}+1\right)^{-1 / 2}.
\end{equation}
Here $F_{\rm peak}= F_{\rm S}/u_0$ is the source flux at event peak and $\tau_{\rm eff} \equiv (t-t_0)/t_{\rm eff}$. The joint data set of Leo and L2 can then be modeled with the parameter set
\begin{equation}
    \bdv{a}=(t_{0,\l2},\,t_{\eff,\l2},\,F_{\rm peak,\l2},\,t_{0,\leo},\,t_{\eff,\leo},\,F_{\rm peak,\leo}) ,
\end{equation}
and the corresponding derivatives are
\begin{equation} \label{eqn:fisher_l2_q}
\frac{\partial F_\l2}{\partial \bdv{a}} = \left(
F_{\rm peak,\l2} Q_{\l2}^{3}\frac{\tau_{\eff,\l2}}{t_{\eff,\l2}}, 
F_{\rm peak,\l2} Q_{\l2}^{3}\frac{\tau_{\eff,\l2}^{2}}{t_{\eff,\l2}}, 
Q_{\l2}, 
0, 
0, 
0
\right);
\end{equation}
\begin{equation} \label{eqn:fisher_leo_q}
\frac{\partial F_\leo}{\partial \bdv{a}} \rightarrow \left(
0,
0,
0,
F_{\rm peak,\leo} Q_{\leo}^{3}\frac{\tau_{\eff,\leo}}{t_{\eff,\leo}}, 
F_{\rm peak,\leo} Q_{\leo}^{3}\frac{\tau_{\eff,\leo}^{2}}{t_{\eff,\leo}}, 
Q_{\leo}
\right).
\end{equation}
The Fisher matrix in Equation~(\ref{fisher}) can then be written as
\begin{equation} \label{eqn:separate_matrix}
\mathcal{F} = \left[\begin{array}{cc}
\mathcal{F}_\l2 & \bdv{0} \\
\bdv{0} &\mathcal{F}_\leo
\end{array}\right],
\end{equation}
where $\mathcal{F}_\l2$ and $\mathcal{F}_\leo$ are both $3\times3$ matrices.

\citet{Penny:2019} performed numerical simulations for the \emph{Roman} microlensing campaign and found that most of the \emph{Roman} microlensing events, especially those with planetary signals, should have $H\lesssim23$, for which the photometric noises were dominated by the source flux (roughly 16--23 mag; see the dashed curve in Figure~\ref{fig:noise-curves}). We can therefore assume that the magnitude uncertainty $\sigma_m \propto F^{-1/2}$. The Fisher matrix given in Equation~(\ref{eqn:separate_matrix}) is then integrable, with
\begin{equation}
\mathcal{F}_{\l2}=\Gamma_{\l2} \left[\begin{array}{ccc}
\frac{C_{1} F_{\text{peak},\l2}}{t_{\eff,\l2}}& 0& 0 \\
0 &\frac{C_{2} F_{\text{peak},\l2}}{t_{\eff,\l2}}& C_{3} \\
0 & C_{3} &\frac{C_{4} t_{\eff,\l2}}{ F_{\text{peak},\l2}}
\end{array}\right]
\end{equation}
and
\begin{equation}
\mathcal{F}_{\leo}=\Gamma_{\leo} \left[\begin{array}{ccc}
\frac{C_{5} F_{\text{peak},\leo}}{t_{\eff,\leo}}& 0& 0 \\
0 &\frac{C_{6} F_{\text{peak},\leo}}{t_{\eff,\leo}}& C_{7} \\
0 & C_{7} &\frac{C_{8} t_{\eff,\leo}}{ F_{\text{peak},\leo}}
\end{array}\right].
\end{equation}
Here $\Gamma_\l2$ and $\Gamma_\leo$ are the observation cadences at L2 and Leo, respectively, and $C_k(k=1, 2, 3 , 4, 5, 6, 7, 8)$ are constants that do not depend on the lensing parameters. Then the covariance matrix $\mathcal{C}_{ij}$ can be given by
\begin{equation}
\mathcal{C}=\left[\begin{array}{cc}
\mathcal{F}_{\l2}^{-1}& 0 \\
0 &\mathcal{F}_{\leo}^{-1}
\end{array}\right] ,
\label{covar}
\end{equation}
with
\begin{equation}
\mathcal{F}_{\l2}^{-1}=\frac{1}{\Gamma_{\l2}} \left[\begin{array}{ccc}
\frac{t_{\eff,\l2}}{C_1 F_{\text{peak},\l2}}& 0& 0 \\
0 &\frac{C_{4} t_{\eff,\l2}}{(C_{2}C_{4}-C_{3}^2)F_{\text{peak},\l2}}&\frac{C_{3}}{C_{3}^2-C_{2}C_{4}}  \\
0 & \frac{C_{3}}{C_{3}^2-C_{2}C_{4}} &\frac{C_{2} F_{\text{peak},\l2}}{(C_{2}C_{4}-C_{3}^2) t_{\eff,\l2}}
\end{array}\right];
\end{equation}
 \begin{equation}
\mathcal{F}_{\leo}^{-1}=\frac{1}{\Gamma_{\leo}} \left[\begin{array}{ccc}
\frac{t_{\eff,\leo}}{C_5 F_{\text{peak},\leo}}& 0& 0 \\
0 &\frac{C_{8} t_{\eff,\leo}}{(C_{6}C_{8}-C_{7}^2)F_{\text{peak},\leo}}&\frac{C_{7}}{C_{7}^2-C_{6}C_{8}}  \\
0 & \frac{C_{7}}{C_{7}^2-C_{6}C_{8}} &\frac{C_{6} F_{\text{peak},\leo}}{(C_{6}C_{8}-C_{7}^2) t_{\eff,\leo}}
\end{array}\right].
\end{equation}
The covariance matrix of the vector $\bdv{\Lambda}$ can be given by
\begin{equation}
\bdv\Sigma_{\bdv\Lambda}= \frac{1}{D_\perp^2}  \left[\begin{array}{cc}
\sigma^2 (\Lambda_\parallel) & 0 \\
0 & \sigma^2 (\Lambda_\perp)
\end{array}\right] ,
\end{equation}
and
\begin{equation}
\left\{ \begin{array}{ll}
\sigma^2 (\Lambda_\parallel) = \frac{1}{C_1 \Gamma_{\l2}} \frac{t_{\eff,\l2}}{F_{\text{peak},\l2}} + \frac{1}{C_5 \Gamma_{\leo}} \frac{t_{\eff,\leo}}{F_{\text{peak},\leo}} \approx \frac{t_{\rm E} u_{0,\leo}^2}{F_{\rm S,\leo}} \left( \frac{1}{C_1 \Gamma_\l2 b} + \frac{1}{C_5 \Gamma_\leo} \right) & \\
\sigma^2 (\Lambda_\perp) = \frac{C_{4}}{(C_{2}C_{4}-C_{3}^2) \Gamma_{\l2}} \frac{t_{\eff,\l2}}{F_{\text{peak},\l2}}+ \frac{C_{8}}{(C_{6}C_{8}-C_{7}^2) \Gamma_{\leo}} \frac{t_{\eff,\leo}}{F_{\text{peak},\leo}} \approx \frac{t_{\rm E} u_{0,\leo}^2}{F_{\rm S,\leo}} \left( \frac{C_4}{(C_2C_4-C_3^2)\Gamma_\l2 b} + \frac{C_8}{C_6C_8-C_7^2)\Gamma_\leo} \right) & 
\end{array} \right. .
\end{equation}
Here we have assumed that $u_{0,\l2} \approx u_{0,\leo}$. With the definition of the $\chi^2$ given by Equation~(\ref{eqn:chi2_definition}), we thus have
\begin{equation}
\chi_{\bdv\Lambda}^{2}=\frac{\Lambda_\parallel^2}{\sigma^{2}(\Lambda_\parallel)}  +\frac{\Lambda_\perp^2}{\sigma^{2}(\Lambda_\perp) } \propto \frac{D_\perp^2 F_{\rm S,\leo}}{t_{\rm E} u_{0,\leo}^2} .
\end{equation}

\bibliographystyle{raa}
\bibliography{my_bib}

\label{lastpage}
\end{CJK*}
\end{document}